# Exploring Qualitative Research Using LLMs

**Muneera Bano, Didar Zowghi, Jon Whittle**

CSIRO's Data61

FirstName.LastName@csiro.au

## Abstract

**Context:** The advent of AI-driven large language models (LLMs), such as ChatGPT 3.5 and GPT-4, have stirred discussions about their role in qualitative research. Some view these as tools to enrich human understanding, while others perceive them as threats to the core values of the discipline.

**Problem:** A significant concern revolves around the disparity between AI-generated classifications and human comprehension, prompting questions about the reliability of AI-derived insights. An "AI echo chamber" could potentially risk the diversity inherent in qualitative research. A minimal overlap between AI and human interpretations amplifies concerns about the fading human element in research.

**Objective:** This study aimed to compare and contrast the comprehension capabilities of humans and LLMs, specifically ChatGPT 3.5 and GPT-4.

**Methodology:** We conducted an experiment with small sample of Alexa app reviews, initially classified by a human analyst. ChatGPT 3.5 and GPT-4 were then asked to classify these reviews and provide the reasoning behind each classification. We compared the results with human classification and reasoning.

**Results:** The research indicated a significant alignment between human and ChatGPT 3.5 classifications in one-third of cases, and a slightly lower alignment with GPT-4 in over a quarter of cases. The two AI models showed a higher alignment, observed in more than half of the instances. However, a consensus across all three methods was seen only in about one-fifth of the classifications. In the comparison of human and LLMs reasoning, it appears that human analysts lean heavily on their individual experiences. As expected, LLMs, on the other hand, base their reasoning on the specific word choices found in app reviews and the functional components of the app itself.

**Conclusion:** Our results highlight the potential for effective human-LLM collaboration, suggesting a synergistic rather than competitive relationship. Researchers must continuously evaluate LLMs' role in their work, thereby fostering a future where AI and humans jointly enrich qualitative research.

## 1. Introduction

Generative AI models, particularly large language models (LLMs) such as OpenAI's ChatGPT 3.5 and GPT-4, are becoming increasingly sophisticated, offering potential applications in a variety of fields (Van Dis et al. 2023). These advanced AI applications have been meticulously designed and trained on vast datasets, allowing them to generate human-like text that can answer questions, write essays, summarize text, and even engage in conversations (Dergaa et al. 2023). The promise they offer is not just in their ability to process information but also in their potential to mimic human-like comprehension and generation of text (Byun, Vasicek, and Seppi 2023).

The transformative influence of these LLMs is being felt across a variety of fields, but perhaps one of the most intriguing applications lies in the domain of qualitative research. Traditionally, qualitative research has always hinged upon the unique human ability to interpret nuances and discern underlying meanings from complex, often ambiguous data. However, the advent of LLMs, with their ability to handle large volumes of data, identify intricate patterns, and generate contextually appropriate responses, has sparked curiosity about their possible roles in qualitative research. The confluence of LLMs and qualitative research offers tantalizing possibilities, but it also raises profound questions. How reliable and valid are AI-generated interpretations compared to those derived from human understanding? What are the implications if the two do not align?

Since the debut of OpenAI's ChatGPT in November 2022, there has been a surge of academic interest in analyzing its potentials across various fields of study. A survey of Scopus[1] revealed that 587 papers reference ChatGPT in their titles or abstracts, with a distribution that spans diverse domains: 247 from medicine, 147 from social sciences, 116 from computer science, and 83 from engineering. Google Scholar[2] lists 7200 articles with ChatGPT mentioned in their titles, with a dominant entries coming from the health and education sectors. This indicates an unexpectedly swift adoption of this AI tool by researchers in the health and social science disciplines. A systematic review (Sallam 2023) of health education using ChatGPT shows 85% of the 60 included records praised the merits of ChatGPT, underlining its effectiveness in improving scientific writing, research versatility, conducting efficient data analysis, generating code, assisting in literature reviews, optimizing workflows, enhancing personalized learning, and bolstering critical thinking skills in problem-based learning, to name a few.

However, employing LLMs in specialized research could potentially introduce issues such as inaccuracies, bias, and plagiarism. Upon tasking ChatGPT with a set of medical research queries related to depression and anxiety disorders, it was observed that the model often produced inaccurate, overblown, or misleading as reported in (Van Dis et al. 2023). These errors could arise from an inadequate representation of relevant articles in ChatGPT's training data, an inability to extract pertinent information, or a failure to distinguish between credible and less credible sources. Evidently, LLMs may not only mirror but potentially amplify human cognitive biases (James Manyika 2019) such as availability, selection, and confirmation biases (Kliegr, Bahník, and Fürnkranz 2021; Bertrand et al. 2022).

The discourse concerning the potential replacement of humans by machines, and the capacities in which this may occur, has already gained significant momentum (Chui, Manyika, and Miremadi 2016; Michel 2020; Prahl and Van Swol 2021). A parallel debate addresses the degree to which machines, particularly artificial intelligence, embody elements of human traits or personhood. Scholarly literature includes investigations into the theoretical creative autonomy attributed to AI poets (Amerika, Kim, and Gallagher 2020), the experiences ostensibly undergone by our smart devices (Akmal and Coulton 2020), and provocative inquiries into the existence of souls within voice assistants (Seymour and Van Kleek 2020). Adding a dramatic dimension to this discourse, a former Google engineer postulated that Google's language models, specifically LaMDA (Thoppilan et al. 2022), possess sentience and are therefore entitled to rights typically reserved for humans (Griffiths 2022).

---

[1] Search was conducted on 14th June 2023 with just one keyword "ChatGPT" to appear in title, abstract of keywords of the published papers.
[2] Search was conducted on 14th June 2023 with just one keyword ChatGPT to appear in title of the published papers.

In response to the doomsday hype of 'LLMs replacing the Human Researcher' (Cuthbertson 2023), our research aims to examine the alignment between human and AI comprehension. We designed an experiment using Schwartz's human values framework (Schwartz 2012). Specifically, we delve into the comparison of LLMs-driven and human classifications of Alexa voice assistant app reviews, as they provide rich and diverse qualitative data. Our goal is to unravel the extent to which LLMs can replicate or align with human understanding and the implications of any misalignment.

The contribution of our research lies in providing much-needed insights, derived primarily from an experiment, into the intersection of AI and qualitative research, a rapidly evolving area with significant implications for the future of the field. By exploring the capabilities and limitations of LLMs in understanding and interpreting qualitative data, we offer a valuable contribution to the ongoing discourse around AI's role in qualitative research.

The organization of this article adheres to the following structure: Section 2 contextualizes the research through an exploration of background information and a review of relevant literature. The subsequent Section 3 describes the design of our exploratory experiment. The results of the investigation are presented in Section 4, which is followed by a discussion in Section 5 that contemplates the repercussions of these results and offers critical insights applicable to qualitative research methodologies. Finally, Section 6 outlines the limitations of our research and Section 7 draws the study to a close by presenting a conclusion and delineating potential avenues for future research.

## 2. Background and Related Work

Large Language Models such as ChatGPT, have generated extensive interest and research across various academic fields in less than a year of its launch. The existing literature on the topic covers several domains, including AI's application in research and academia, its role in education, its performance in specific tasks, and its use in particular sectors like library information centers and medical education.

Numerous studies have explored the capabilities and limitations of AI in research and academia. Tafferner et al. (2023) analyzed the use of ChatGPT in the field of electronics research and development, specifically in applied sensors in embedded electronic systems. Their findings showed that the AI could make appropriate recommendations but also cautioned against occasional errors and fabricated citations.

Kooli (2023) delved into the ethical aspects of AI and chatbots in academia, highlighting the need for adaptation to their evolving landscape. Echoing this sentiment, Qasem (2023) explored the potential risk of plagiarism that could stem from the misuse of AI tools like ChatGPT. To balance the benefits and potential misuse, Burger et al. (2023) developed guidelines for employing AI in scientific research processes, emphasizing both the advantages of objectivity and repeatability and the limitations rooted in the architecture of general-purpose models.

The role of AI in education is another pivotal theme in the literature. Wardat et al. (2023) investigated stakeholder perspectives on using ChatGPT in teaching mathematics, identifying potential benefits and limitations. Similarly, Yan (2023) explored the use of ChatGPT in language instruction, pointing to its potential but also raising concerns about academic honesty and educational equity Jeon and Lee (2023) examined the relationship between teachers and AI, identifying several roles for both and emphasizing the continued importance of teacher's pedagogical expertise. In a broader study of public discourse and user experiences, Tlili et al. (2023) highlighted a generally positive perception of AI in education but also raised several ethical concerns.

A strand of research has also evaluated AI's performance in specific tasks traditionally conducted by humans. Byun, Vasicek, and Seppi (2023) showed that AI can conduct qualitative analysis and generate nuanced results comparable to those of human researchers. In another task-specific study, (Gilson et al. 2023) demonstrated that ChatGPT could answer medical examination questions at a level similar to a third-year medical student, underscoring its potential as an educational tool.

Research has explored the use of ChatGPT in specific sectors. Panda and Kaur (2023) investigated the viability of deploying ChatGPT-based chatbot systems in libraries and information centers, concluding that the AI could provide more personalized responses and improve user experience. Similarly, Gilson et al. (2023) indicated the potential of ChatGPT as an interactive medical education tool, further expanding the potential application areas of AI in different sectors.

Although existing literature has extensively covered AI's impact in various domains, several gaps remain. Notably, a lack of qualitative research comparing human reasoning against LLMs is evident. Burger et al. (2023) and Byun et al. (2023) have made an initial foray into this area, demonstrating that ChatGPT can perform certain research tasks traditionally undertaken by human researchers, producing complex and nuanced analyses of qualitative data with results arguably comparable to human-generated outputs. Despite these promising findings, these studies do not investigate AI and human reasoning within the qualitative research context.

## 3. Experiment Design

The aim of our research was to compare and contrast human comprehension with that of two LLMs: ChatGPT 3.5 and GPT-4, specifically within the context of qualitative research. We sought to understand the depth to which these LLMs could analyze and provide reasoning for their judgment.

To design and conduct our experiment (see Figure 1), we leveraged the framework of Schwartz's theory of human values, a well-regarded model that encapsulates ten basic universal values present across cultures (Schwartz 2012). These include power, achievement, hedonism, stimulation, self-direction, universalism, benevolence, tradition, conformity, and security. This conceptual schema enabled us to perform a comparative analysis between human and AI reasoning within a structured and widely accepted paradigm of human values.

We designed an experiment to explore through a case of the Amazon Alexa voice assistant app's reviews. These reviews provide a rich source of qualitative data, with users expressing their opinions, perceptions, and values implicitly or explicitly in their feedback (Shams et al. 2021). We randomly selected a sample of 18 Alexa app reviews from a set that had previously been classified by a human analyst according to Schwartz's human values (Shams et al. 2023). This study created a benchmark for our comparison with the classifications generated by the LLMs. Shams et al. (2013) in their study were aiming to conduct an empirical analysis of user feedback for Amazon's Alexa app to identify a set of essential human values and validate them as requirements for AI systems within distinct usage contexts, a technique that could potentially be extrapolated to other AI platforms.

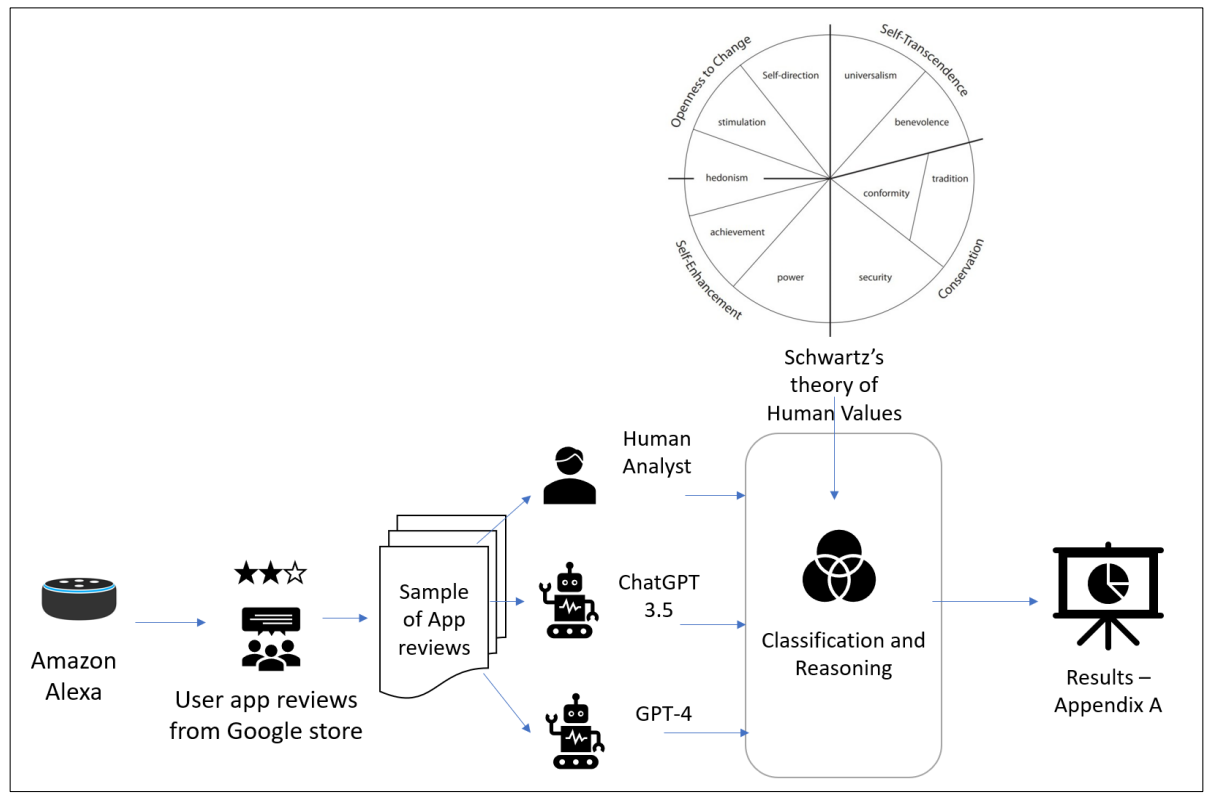

*Figure 1. Experiment Design*

Though a randomized selection of 18 reviews represents a limited sample, our primary objective was not to focus on the sample size but to explore the variation in responses between the human analyst and LLMs. Our priority was mainly on the 'why' aspect of all classifications.

Our experiment involved prompting ChatGPT 3.5 and GPT-4 to generate their classifications for the same reviews. We were interested not only in their classification outcomes but also in their rationale for each categorization. This design aimed to gauge the LLMs' depth and their capability to reason within the context of Schwartz's human values framework as compared to human comprehension.

Designing appropriate prompts for LLMs is a critical process (White et al. 2023) and it can significantly influence the outcome of any LLMs' analysis. The composition and specificity of prompts can guide the models' analysis and processing of the task, thus affecting the results. A well-structured, clear, and contextually rich prompt helps the LLMs focus on the essential aspects of the task, reducing the likelihood of errors or misinterpretations. For every individual app review, we used the same prompt as below for both ChatGPT 3.5 and GPT4:

*Following is an app review from a user of Amazon Alexa. Analyse the review text and classify it against Schwartz's theory for Human Values, both main and sub values. Provide your reason on why you classified it against that value.*

Our prompt design considered these three elements (structure, clarity, and context). Firstly, the prompt is clear as it explicitly outlines the task at hand, namely the analysis of an Amazon Alexa user review. Secondly, it displays structure, sequentially detailing each step to be undertaken, beginning with the review analysis, followed by classification against Schwartz's theory of Human Values, and concluding with an explanation for the chosen classification. Lastly, the prompt provides the context; it not only specifies the source of the review (Amazon Alexa) but also guides the model to employ a particular theoretical framework (Schwartz's theory of Human Values). Such specificity allows the model to tune its responses based on the understanding of the context provided, including both main and sub-values, thereby facilitating a more nuanced analysis.

The first author conducted the entire experiment and carried out the conversation with LLMs. The complete results of all the responses from both LLMs and their comparison with human analysis are presented in Appendix A. To triangulate the results obtained from the comparisons and further discuss the findings and insights, the second and third authors then conducted an independent review of the results obtained from the human and LLMs to form an opinion about the reasonability of the classifications.

## 4. Results

Based on the analysis of the classifications (see Appendix A), several intriguing observations were noted. In 33% (6 out of 18) of the instances, there was a close alignment between human and ChatGPT 3.5 classifications. Similarly, in 27% (5 out of 18) of cases, the classifications of humans and GPT-4 exhibited close correspondence. Interestingly, the agreement between ChatGPT 3.5 and GPT-4 classifications was higher, found in 56% (10 out of 18) of the cases. Moreover, only in 22% (4 out of 18) of instances, there was a unanimous agreement amongst all three - human, ChatGPT 3.5, and GPT-4.

The divergences found among the classifications trigger compelling questions about the reliability of the results generated by LLMs. The inconsistencies among the insights derived from LLMs and human interpretations lead to speculation about the capability of LLMs in fully appreciating and navigating the intricacies of human language and contextual nuances. This view is especially prevalent among qualitative researchers who consider these discrepancies as a warning that LLMs might not be adequately equipped.

The notable alignment between ChatGPT 3.5 and GPT-4, reaching 56%, highlights concerns about the propensity of LLMs to inadvertently foster a singular perspective, which may be creating an AI echo chamber in qualitative research. This could potentially lead to the obliteration of diversified thought, an element crucial to enrich and deepening the analysis in qualitative research.

Another facet that emerged from our analysis was that in some instances classifications made by ChatGPT 3.5 and GPT-4 appeared to be more logical and reasonable. This was determined by the triangulation conducted by the second and third authors comparing LLMs classifications to the humans. For example , in one of the instances in Appendix A, the human analyst classifies the review: "I'd enjoy and find this app very useful if it did WHAT it was supposed to WHEN it was supposed to" as "Benevolence" and "Loyalty". While ChatGPT 3.5 classifies it as "Achievement" and "Competence", and GPT4 mentions "Achievement" and "Capability". The two peer reviewers considered answers from LLMs to be more logical and reasonable than that of humans. This could suggest that LLMs can offer a fresh, alternative perspective that might not have been identified by human researchers.

Table 1 showcases three examples where human analysts and LLMs, diverge significantly in their respective reasoning. Human analysts appear to prioritize emotional and personal facets of the situation, while LLMs frequently underscore functional or skill-oriented aspects. For instance, Example 1 displays a human analyst emphasizing user curiosity, while LLMs consider Alexa's capabilities. Moreover, the human analyst seems to provide interpretations imbued with more contextual sensitivity, while LLMs occasionally overlook subtle nuances embedded within the text.

|  | Example 1<br>Divergent Categorizations | Example 2<br>Conceptual Agreement | Example 3<br>Consistent Categorizations |
|---|---|---|---|
| **App Review** | I've asked difficult things, but every time she's fantastic at answering. | Alexa shares data (utterances, notifications) between household accounts violating the users' privacy. | Why do the new gen Alexa not have the replies as the original? Are the new programmers so selfish and disrespectful that they don't know how to say thank you or you are welcome, since when did politeness come at a cost? |
| **Human Categorization** | Self-direction and Curiosity | Self-direction and Privacy | Conformity and Politeness |
| **ChatGPT 3.5 Categorization** | Achievement and Competence | Security and Privacy | Conformity and Politeness |
| **GPT-4 Categorization** | Universalism and Knowledge | Security and Privacy | Conformity and Politeness |
| **Human vs. AI reasoning** | Human analysts tend to focus on the emotional and personal aspects of a situation, while AI models may focus more on the functional or skill-based aspects. In this case, the human analyst identified the user's curiosity, while both AI models concentrated on Alexa's abilities. | The human analyst combines "Self-direction" with "Privacy," emphasizing the individual's autonomy and control over their own data. In contrast, both AI models focus specifically on the "Security and Privacy" aspect, highlighting the broader implications of data sharing. | The human analyst points out the specific polite responses that are missing from the new generation of Alexa, such as "thanks" and "welcome." In contrast, both AI models focus on the general criticism of Alexa's lack of polite responses, without specifying the exact phrases. |
| **Insights** | **Context sensitivity:** AI models may not always capture the context or nuances of a text as accurately as a human analyst. Training AI models on a broader range of examples and refining their understanding of context could help improve their categorizations.<br>**Different perspectives:** The human analyst and AI models approached the categorization from different perspectives, showing that multiple interpretations can be valid. This highlights the importance of considering diverse viewpoints in qualitative analysis and recognizing that there may not always be a single "correct" answer. | **Interpretation of Privacy:** The human analyst explicitly mentions the lack of consent, which indicates a violation of the user's autonomy. In comparison, the AI models focus on the concern about sharing personal data, but they do not emphasize the lack of consent as the human analyst does. | **Interpretation of the issue:** The human analyst attributes the issue to Alexa being "impolite," while both AI models focus on the user's disapproval or criticism. The human analyst's reasoning seems to take a more personal stance, whereas the AI models present a more neutral description of the user's concern. |

*Table 1. Comparison of three scenarios of Human vs AI agreement*

These disparities in reasoning provoke a contentious discourse about the aptness of LLMs in qualitative research. Some researchers assert that LLMs, given their current technological stature, are incapable of completely comprehending the profound complexities of human emotions and experiences (Bender et al. 2021; Alkaissi and McFarlane 2023; Rudolph, Tan, and Tan 2023). Consequently, their use in qualitative analysis should be treated with caution. The argument furthers that the LLMs missing context sensitivity and focus on functional aspects could lead to flawed or incomplete conclusions. But the prompts developed by humans need to provide a rich context in order to address this issue.

Contrastingly, advocates of AI-assisted qualitative analysis propose that LLMs can furnish invaluable insights and complementary viewpoints, aiding researchers in achieving a more all-encompassing understanding of the data (Dwivedi et al. 2023). The researchers in favour of LLMs further posit that with the consistent evolution and enhancement of AI, a synergistic approach combining human acumen and AI capabilities can lead to more robust analysis.

This ongoing discussion brings forth crucial questions for qualitative researchers concerning the degree of their reliance on LLMs in their work. While LLMs hold the potential to transform qualitative research by delivering additional perspectives and insights, it is imperative for researchers to also acknowledge their limitations and maintain a keen awareness of the humanistic elements inherent to qualitative research.

## 5. Discussion

*AI Doomsday*

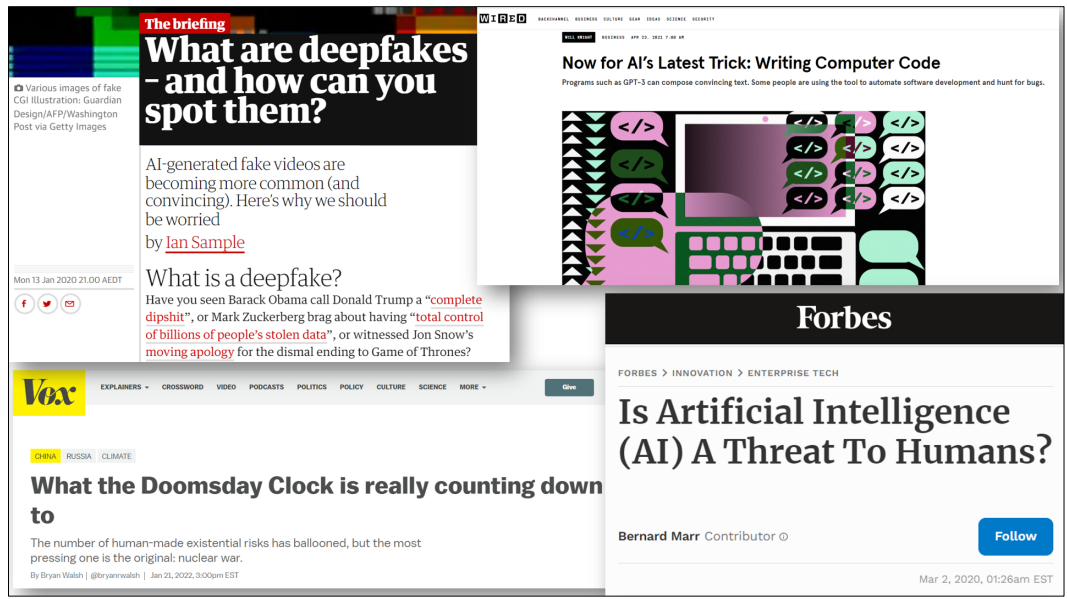

*Figure 2. Media amplification of AI Doomsday fears*

The escalating discourse on the potential risks of AI and LLMs, amplified by recent media reports (Figure 2), is leading to a growing unease among various professional communities. They are coming to terms with the stark reality that AI might soon eclipse their roles and replace them in their jobs. This existential dread has been underscored by developments such as the AI Doomsday Clock[3] inching closer to midnight, symbolizing the perceived imminent danger of a catastrophic AI disaster.

Findings from our exploratory experiment, coupled with an overview of existing research and an understanding of capabilities of LLMs, do not support a doomsday scenario for qualitative researchers. Contrary to pervasive fears, the reality we've discerned suggests a future where human researchers and LLMs can coexist and contribute complementarily to the field of qualitative research.

*AI and Humans*

Despite the considerable potential of LLMs in qualitative research, the indispensable role of the human researcher for verifying the validity and reliability of the results remains critical. LLMs, while robust and efficient, exhibit limitations in their understanding of complex human experiences, contexts, and semantics, occasionally leading to the generation of inaccurate or invented information, a phenomenon known as 'hallucinations' (Rudolph, Tan, and Tan 2023; Alkaissi and McFarlane 2023). These hallucinations can misdirect the interpretation of research results, compromise validity, and introduce unintentional bias or error. Therefore, the human researchers' involvement becomes vital in scrutinizing, verifying, and interpreting the results generated by LLMs, ensuring that the outcomes are consistent with the actual context and preserving the integrity of the research. Furthermore, the human researcher's expertise and critical thinking are required to continually calibrate these models and improve their comprehension over time, helping in enhancing their capabilities while minimizing potential drawbacks.

---

[3] https://www.vox.com/22893594/doomsday-clock-nuclear-war-climate-change-risk

LLMs are poised to redefine the interplay between AI and human involvement in the research process. When it comes to inductive reasoning and open-ended data collection, LLMs are capable of deriving insights from unstructured data without predetermined hypotheses and continuously collecting and analyzing massive amounts of data from diverse sources. However, while these capabilities can expedite the research process, the question remains whether LLMs can truly replicate the intuitive reasoning processes and interpretive nuances inherent to human researchers. Similarly, while LLMs can process large amounts of qualitative data collected in naturalistic settings, the nuanced understanding, cultural sensitivity, and context-awareness that human researchers bring to these settings are unlikely replicable by LLMs in their current state.

Polanyi's concept of 'tacit knowledge' (Collins 2005) which is also the 'implicit' component of Nonaka's SECI model (Li and Gao 2003) underscores the unique human ability to perform certain tasks in unexpected and inexplicable ways. This inherent capability, however, may not be explicitly replicated or comprehended by LLMs, due to the unpredictable nature of such knowledge that is often grounded in personal experience and intuition.

A further manifestation of Human-LLM collaborative research could involve delineating distinct roles for each entity to optimize the research process. Here, LLMs could function as 'inter-rater reliability testers' (Armstrong et al. 1997), contributing to the research conducted by human analysts, while the human participants would be responsible for the verification of the information and analytical results generated by the LLMs. This iterative process, involving reciprocal roles, has the potential to yield more robust and efficient research outcomes, underscoring the mutual enrichment of human insight and machine efficiency.

*Stochastic Parrots for Qualitative Research*

LLMs like GPT-4 and ChatGPT have demonstrated remarkable capacity to generate human-like text, understand context, and interact dynamically with users. Their potential, however, should not overshadow the challenges they pose, especially concerning the interpretation of meanings in qualitative research. A significant advantage of these models lies in their ability to process and analyze vast amounts of data quickly and relatively accurately, providing a broad view of patterns and trends that could otherwise be missed.

Nonetheless, Bender et al. present cogent arguments about the risks associated with these models, primarily centered around their training on massive and diverse text datasets (Bender et al. 2021). This training can result in the replication and amplification of biases present in the data, leading to potentially harmful outputs. Additionally, the text generation process of LLMs remains fundamentally opaque, raising questions about transparency and interpretability.

Despite LLMs' adeptness at generating linguistically coherent responses, they do not genuinely comprehend the meanings, nuances, and deeper implications of words and phrases. While humans possess a holistic understanding of language, encompassing cultural, emotional, historical, and symbolic dimensions, LLMs can only provide approximations based on learned patterns. They may miss out on the rich tapestry of meanings a human researcher could decipher.

To mitigate the risks associated with LLMs, (Bender et al. 2021) propose several steps, including (a) reducing model size, (b) increasing transparency, and (c) establishing ethical guidelines for their use. Smaller, more controlled models could potentially minimize harm, while greater transparency could facilitate a better understanding of the mechanisms behind their text generation. Ethical guidelines would also establish a framework for responsible and equitable use of these models.

The determination of an appropriate size for LLMs, a balance between the model's complexity and its predictive accuracy, is best achieved through a collaboration of machine learning experts, ethicists, and domain-specific experts. As for the selection of ethical guidelines governing LLMs use should be context-dependent and reflective of the values and perspectives of a diverse range of stakeholders (Zowghi and da Rimini 2023). This selection process necessitates an inclusive approach, possibly involving a blend of established ethical frameworks tailored to the specifics of the AI system and its deployment (Sanderson et al. 2023). The ethical guidelines established by a diverse and inclusive committee of stakeholders need to be periodically reviewed and updated to align with evolving societal norms and technological advancements.

The use of LLMs in qualitative research also introduces a new set of ethical considerations. Concerns around privacy, data misuse, and the risk of perpetuating existing biases in the data they are trained on are prevalent. Additionally, the advent of LLMs in the academic sphere raises questions about intellectual property rights and authorship. In this changing landscape, the role of the human researcher may shift towards orchestrating the research process, ensuring ethical compliance, and interpreting and contextualizing the findings generated by LLMs. As technology continues to advance, the importance of critical reflection on these shifts and their implications will grow.

*Evolution of Qualitative Research*

LLMs have the potential to significantly impact data analysis in qualitative research, as they can speed up the process and handle larger datasets than humans can feasibly manage. For example, a study by (Byun, Vasicek, and Seppi 2023) demonstrated that AI is capable of conducting qualitative analysis and generating nuanced results. However, such studies often do not delve into the reasoning behind AI vs. human interpretation, which could significantly impact the findings. In addition, there is the question of whether LLMs can truly understand and articulate the symbolic and cultural nuances that underpin human behavior, elements that are paramount to the work of prominent anthropologists and sociologists, such as Malinowski's participatory observation (Malinowski 1929), Weber's concept of verstehen, or empathetic understanding (Weber 1949), and Geertz's interpretation of culture (Geertz 1973).

The application of LLMs could potentially enhance the efficiency of established qualitative methodologies such as Grounded Theory (Charmaz 2014; Glaser, Strauss, and Strutzel 1968), Interpretive Interactionism (Denzin 2001), and Narrative Analysis (Franzosi 1998), particularly in terms of initial data analysis. However, these methodologies were developed with the understanding that the researcher's empathy, interpretation, and contextual understanding are integral to the process. As such, it is unlikely that the essential humanistic aspects of these approaches can be fully replaced by LLMs, indicating a shift rather than an absolute transformation in these methodologies (Dwivedi et al. 2023).

## 6. Limitations

While our study provides valuable insights into the utilization of LLMs in qualitative research, these findings are inevitably influenced by our own areas of expertise and the specific experimental design we employed. The research is also constrained by two primary limitations. Firstly, the small sample size we chose for the study, although increasing the sample size could have altered the statistical outcomes. However, our primary interest lay not in large-scale data analysis, but in exploring the reasoning patterns of human analysts and LLMs during the classification of app reviews. Secondly, our selection of app reviews for manual classification was performed by a single human analyst. Future research could benefit from comparing the levels of agreement and discrepancies between multiple human analysts' classifications and contrasting these with the outputs of various LLMs, thereby broadening the scope and depth of the experiment.

## 7. Conclusion and Future Work

The insights obtained from our experiment underscore the significance of careful considerations regarding the use of AI models play in qualitative research. The modest alignment between human and AI classifications, coupled with the comparatively higher concordance between the AI models, illuminates the complex dynamics at play when incorporating AI into qualitative analysis. Our findings accentuate that, despite the promise of AI for augmenting analysis, the unique human touch—an element intrinsic to qualitative research—cannot be disregarded. This essential human element, embedded in understanding and interpreting context, remains a critical factor (for now) in maintaining the richness and depth of qualitative investigations.

The considerable variations highlighted between human and AI comprehension in this study encourage further exploration in the realm of AI integration into qualitative research. Future work could delve deeper into understanding the basis for such disparities, thereby refining the synergistic interplay between AI and human analysis. Furthermore, a close examination of the AI echo chamber effect on diversity in qualitative research is warranted. Moreover, investigating how to leverage the different perspectives offered by AI, while keeping the human touch intact, could lead to more comprehensive and nuanced insights. Lastly, addressing the ethical implications of AI usage in qualitative research, especially considering AI's limitations and potential for biases, will form a critical part of future studies. As we venture further into this new era of AI-assisted research, it is imperative to navigate these challenges to harness the full potential of this technological advancement in a responsible and ethical manner.

# Appendix A – App Review Analysis

| Alexa App review | Human Analysis | | | ChatGPT 3.5 | | | GPT-4 | | |
|---|---|---|---|---|---|---|---|---|---|
| | Main Value | Individual Value | Reason | Main Value | Individual Value | Reason | Main Value | Individual Value | Reason |
| I've asked difficult things but every time she's fantastic at answering | Self-Direction | Curiosity | Alexa was able to meet the user's curiosity by answering all questions. | Achievement | Competence | The user values Alexa's ability to answer difficult questions. | Universalism | Knowledge | Appreciation for Alexa's ability to answer difficult questions |
| Adding photos would be nice if you could rotate photos | Self-Direction | Creativity | User expects creativity in editing photos by Alexa | Self-direction | Creativity | The user would like a new feature added to the app. | Hedonism | Pleasure | Desire for added functionality (rotating photos) to improve experience |
| Doesn't listen to what you say properly has a mind of its own | Self-Direction | Freedom | Alexa violated user's freedom of choice. | Conformity | Restraint | The user criticizes Alexa for not listening properly. | Power | Control | Frustration with Alexa not listening properly and seeming to have a mind of its own |
| Alexa shares data (utterances, notifications) between household accounts violating the users privacy | Self-Direction | Privacy | Clearly violating privacy by sharing data without consent. | Security | Privacy | The user is concerned about Alexa sharing personal data. | Security | Privacy | Concerns about privacy and data sharing between household accounts |
| It doesn't play some songs that I ask. A popular song, for example, would be "Easy on me", by Adele. It'll just choose one of her songs at random but never ever that one | Self-Direction | Choosing own goals | Alexa doesn't listen to the users, therefore, the users are unable to choose their own goals through Alexa. | Stimulation | Novelty | The user is disappointed that Alexa cannot play a specific song. | Hedonism | Pleasure | Disappointment with Alexa not playing specific requested songs |
| Alexa helps in many ways and plays my favorite play list and gives me updates on musicians new albums | Hedonism | Enjoying Life | Alexa is able to entertain users with music and many ways, which confirms users' enjoyment. | Hedonism | Sensuous gratification | The user enjoys using Alexa to play music and receive updates. | Hedonism | Pleasure | Enjoyment from Alexa's assistance, music updates, and playing favorite playlists |
| My younger sibling has down-syndrome and can't say words as clearly as others. When I tried to do the Voice ID it didn't work and it said that No one has said anything in a while, although my sibling had said what it said your supposed to say and it didn't register it, He's very stressed and upset about this and so am I. | Universalism | Equality | Alexa did not ensure equality by not recognizing the voice of a person with down-syndrome. | Benevolence | Helping others | The user is frustrated that Alexa does not recognize their sibling's voice. | Benevolence | Care | Concern for sibling's experience and the lack of accessibility for those with speech difficulties |
| Alexa suddenly at 3am, 4am or whenever wakes me up saying: "connected to..." OR it makes a two notes sound | Universalism | Peace | Nobody expects someone to destroy his/her peaceful sleep at midnight, what Alexa has done successfully! :D | Security | Safety | The user is woken up unexpectedly by Alexa at odd hours. | Security | Safety | Disturbance caused by Alexa's unexpected noises at night |
| Greatest conversation and best way to get informed on Wikipedia information, historical events, everyday events and weather info | Universalism | Wisdom | Alexa is giving much useful information which might enhance the wisdom of its users. | Universalism | Understanding and appreciation | The user values Alexa's ability to provide information and updates. | Universalism | Knowledge | Appreciation for the wealth of information provided by Alexa |
| I use alexa basically to be productive, and I add a lot of tasks and reminders. | Conformity | Self-discipline | Alexa is helping to be organized and disciplined by adding to do list and reminders. | Achievement | Productivity | The user values Alexa for task and reminder management. | Achievement | Capability | Utilization of Alexa for productivity, tasks, and reminders |
| Why do the new gen alexa not have the replies as the original. Are the new programmers so selfish and disrespectful that they don't know how to say thank you or you are welcome, since when did politeness come at a cost | Conformity | Politeness | Alexa found to be impolite, does not even say "thanks" or "welcome". | Conformity | Politeness | The user criticizes Alexa for lacking polite responses. | Conformity | Politeness | Disapproval of Alexa's lack of politeness in newer versions |
| Alexa doesn't listen to me | Conformity | Obedience | Alexa is disobedient as it doesn't listen to its users. | Conformity | Restraint | The user is frustrated that Alexa does not listen to them. | Power | Control | Frustration with Alexa not listening to user's commands |
| Alexa is a privacy and security nightmare from stem to stern | Security | Security | The user is clearly complaining about security threat. | Security | Privacy | The user expresses concerns about Alexa's privacy and security. | Security | Privacy, Safety | Concerns about privacy and security issues with Alexa |
| She has not been responding to my commands. Commands I give everyday, she all of the sudden either doesn't respond to or doesn't understand what I'm saying or just ignores | Power | Authority | Alexa doesn't listen to its users and does whatever she wants. Therefore, users do not have any authority over Alexa to make her do a specific task. | Conformity | Restraint | The user is frustrated that Alexa does not respond to commands. | Power | Control | Frustration with Alexa not responding or understanding everyday commands |
| Only speak English and Hindi but we need some of our language (Tamil, Telugu, Malayalam and more | Tradition | Respect for tradition | Users expect Alexa to understand traditional (local) language. | Universalism | Diversity | The user would like Alexa to support more languages. | Universalism | Understanding | Desire for more language options to cater to diverse users |
| Very frustrating that I have been told if I downloaded the app, I | Benevolence | Honesty | Alexa did not do what it supposed to do. Users | Security | Safety | The user is disappointed that Alexa does not | Achievement | Capability | Frustration with the limitation of connecting only |

| | | | | | | | | |
|---|---|---|---|---|---|---|---|---|
| could have more than one phone connected to my Alexa at once. This is NOT true. | | | expect this much of honesty from Alexa, which it clearly violated. | | | allow multiple phones to connect. | | | one phone to Alexa |
| I'd enjoy and find this app very useful if it did WHAT it was supposed to WHEN it was supposed to | Benevolence | Loyalty | Alexa is not loyal in its activity, it does not do what it's supposed to do | Achievement | Competence | The user would like Alexa to perform as expected. | Achievement | Capability | Expectation for the app to function as intended for a more useful experience |
| Not brilliant. Seems to be a bit like my dog, suffers with selective hearing | Achievement | Intelligence | User is clearly complaining about Alexa's lack of intelligence due to its limitation of understanding (hearing). | Conformity | Restraint | The user is frustrated that Alexa does not respond properly. | Power | Control | Comparison to a dog with selective hearing, suggesting lack of control over Alexa's responsiveness |